\documentstyle[12pt,aasms4]{article}

\newcommand{\hmp}{h^{-1}{\rm Mpc}}

\newcommand{\be}{\begin{equation}}
\newcommand{\ee}{\end{equation}}
\newcommand{\bea}{\begin{eqnarray}}
\newcommand{\eea}{\end{eqnarray}}

\newcommand{\bef}{\begin{figure}}
\newcommand{\eef}{\end{figure}}

\def\spose#1{\hbox to 0pt{#1\hss}} 
\def\ltapprox{\mathrel{\spose{\lower 3pt\hbox{$\mathchar"218$}} 
 \raise 2.0pt\hbox{$\mathchar"13C$}}} 
\def\gtapprox{\mathrel{\spose{\lower 3pt\hbox{$\mathchar"218$}} 
 \raise 2.0pt\hbox{$\mathchar"13E$}}} 
\def\inapprox{\mathrel{\spose{\lower 3pt\hbox{$\mathchar"218$}} 
 \raise 2.0pt\hbox{$\mathchar"232$}}} 
 
 \slugcomment{Submitted to Astrophys. J. Letters}

\lefthead{Joyce \& Sylos Labini}
\righthead{E}

\begin{document} 
 
\title{Luminosity density estimation from redshift surveys 
and the mass density of the Universe}
\author{Michael Joyce \altaffilmark{1} \, and \,
Francesco Sylos Labini \altaffilmark{2,3}}

\altaffiltext{1}{LPT, Universit\'e Paris IX, B\^atiment 211, F-91405 
Orsay, France} 

\altaffiltext{2}{D\'epartement de Physique Th\'eorique, 
Universit\'e de Gen\`eve,
24 Quai Ernest Ansermet, CH-1211 Gen\`eve 4, Switzerland}

\altaffiltext{3}{INFM Sezione Roma1,        
		      Dip. di Fisica, Universit\'a ``La Sapienza'', 
		      P.le A. Moro, 2,  
        	      I-00185 Roma, Italy. }

\begin{abstract} 

In most direct estimates of the mass density (visible or dark) of the 
Universe, a central input parameter is the luminosity
density of the Universe. Here we consider the measurement
of this luminosity density from red-shift surveys, as a function
of the yet undetermined characteristic scale $R_H$ at which the 
spatial distribution of visible matter tends to a well defined
homogeneity. Making the canonical assumption that the cluster mass 
to luminosity ratio ${\cal M}/{\cal L}$ is the universal one, we can 
estimate the total mass density as a function 
$\Omega_m(R_H,{\cal M}/{\cal L})$. Taking the highest estimated cluster 
value ${\cal M}/{\cal L} \approx 300 h M_{\odot}/L_{\odot}$ 
and a conservative lower limit $R_H \gtapprox 20 \hmp$,
we obtain the upper bound $\Omega_m \ltapprox 0.1$ . We note that 
for values of the homogeneity scale $R_H$ in the range  
$R_H \approx (90 \pm 45) h$Mpc, the value of 
$\Omega_m$ may be compatible with the nucleosynthesis inferred 
density in baryons. 
\end{abstract}


\newpage

\setcounter{footnote}{0}
While the density of visible matter can be directly inferred
from red-shift surveys, that of dark matter is necessarily 
only more indirectly accessible. The currently most popular methods
for determining the latter make use of estimates of the total 
mass of clusters, which are stated as 
estimates for the mass to luminosity ratio for clusters (see
Bahcall 1999 for a review and Hradecky et al. 2000 for recent
determinations in groups and clusters of galaxies).
Assuming that this ratio is representative of the global ratio of mass 
to luminosity one can infer, {\it given the luminosity 
density $\phi_*$ of the Universe}, 
the corresponding value for the mass density $\Omega_m$. Almost
invariably in the literature this luminosity density is taken 
from an analysis reported in Efstathiou et al. (1988), which derives
a value $\phi_*$ primarily by fitting the normalization
of number counts from the origin as a function of apparent magnitude.
The intrinsic problems of inferring what is a three dimensional property 
from such two dimensional (projected) measures have been discussed in detail
elsewhere (e.g.  Sylos Labini, Montuori \& 
Pietronero 1998 -hereafter SLMP98). We limit ourselves
to noting here that this procedure of normalization is highly
sensitive to the (a priori unknown) corrections which are 
applied to the data, which
without such corrections do not even show a counts' slope consistent 
with the homogeneity assumption (i.e. $\alpha =0.6$) 
in any range of apparent magnitude. 
Further, despite the quite precise value (error of 
$20 \%$) derived in Efstathiou et al., the values of 
$\phi_*$ found in different surveys
in the paper {\it{vary by a factor of four}}. The authors  
note that ``the dispersion amongst the estimates is
extraordinarily large'', but do not provide any clear explanation
of why this is so and how they arrive at such a small error in
their final (averaged) estimate for $\phi_*$. It is our view
that such variations are intrinsic to the data, corresponding
to the fact that each result is normalized to samples of different
size and geometry. In any case it is clear that the most reliable
way of estimating this three dimensional property of the data is
directly from three dimensional data, and it is this which we do
here. 

Before entering the details of data analysis, let us discuss 
important methodological point which is at the centre of our treatment:
instead of assuming homogeneity in the data sets we analyse, we use 
statistical methods which do not depend on the presence or absence of
such homogeneity. Here this means that instead of assuming that $\phi^*$
is well defined a priori, we evaluate a more general quantity which 
will correspond to it in the case that the distribution is homogeneous. 
Homogeneity at large scales in the distribution of matter is a 
central assumption of standard cosmological models and  of the 
statistical tools usually used to analyze the data.
While two dimensional (angular) maps of galaxies initially 
provided clear support for the supposition of homogeneity at relatively 
small scales, three dimensional red-shift surveys revealed 
unsuspected structure at much larger scales. The existence of
such structures (in particular voids) is incompatible
with well defined homogeneity below these scales, and observationally 
the actual scale characterizing homogeneity in matter, and indeed 
the existence of such a scale at all, has become a subject of 
considerable debate (e.g SLMP98 and Wu, Lahav \& Rees 1999). Standard
characterizations of the three dimensional galaxy and cluster
red-shift data, which simply assume homogeneity at the largest
scale probed, give rise to an ever-growing range of characteristic
scales (``correlation lengths'' of different classes of objects), 
while an analysis of the same surveys with methods which
do not assume such homogeneity leads to an interpretation in which
these scales are sample-dependent characterizations of a distribution
with an underlying simple scale invariance,
a cut-off from which to homogeneity
has,  it is argued,  not yet been detected with any statistical 
significance (see e.g. Joyce, Montuori \& Sylos Labini 1999 -
hereafter JMSL99). Our aim in this paper is not to address 
these issues which are discussed in detail elsewhere, but to 
show how crucial input parameters to standard 
cosmological models depend on what are, at the very least, 
important observational uncertainties concerning the 
distribution of visible matter. In particular we consider here 
the total mass density of the Universe $\Omega_m$, but the same 
kind of analysis can be easily applied to other important parameters
(for example, to the amplitude of the matter power spectrum). 
We express our results in terms of a well defined homogeneity 
scale to be determined from red-shift surveys and place 
bounds on $\Omega_m$ corresponding to conservative current lower 
bounds on this scale.

We first consider the properties of the spatial distribution
of visible matter and the characterization of the tendency (if any)
to homogeneity. At small scales, at least up to $\sim 20 h^{-1}$ Mpc,
it is widely agreed that the galaxy distribution shows fractal 
behavior. Deviation away from this behavior towards the expected
homogeneity is most easily identified using a very 
simple two point statistic, the average conditional density 
\be 
\label{gamma} 
\Gamma(r) = 
\left \langle \frac{1}{S(r)} \frac{dN(<r)}{dr} \right \rangle_p 
\ee 
where $dN(<r)$ is the number of points in a shell 
of thickness $dr$ at distance $r$ from an {\it occupied point} and 
$S(r)dr$ is the volume of the shell. The symbol $\langle ... \rangle_p$ 
indicates that the quantity is a conditional one
(for a discussion see Gabrielli \& Sylos Labini 2001), the average
being performed over occupied points. This statistic is simply an
unnormalized form  of the standard two point correlation functions 
$\xi(r)$ (SLMP98).
While homogeneity corresponds to the convergence of $\Gamma$ 
to a fixed value as a function of distance, scale-invariance is 
indicated by the 
continuation of a simple power law behavior. 
Only in the former case does a real average density 
exist. 
The small scale fractal behaviour observed in red-shift catalogues 
corresponds to the behaviour $\Gamma(r) = Ar^{D-3}$ 
(with $A$ a constant); detecting homogeneity corresponds to making
an estimate of the asymptotic value of the density $\Gamma_{\infty} > 0$;
we then define $AR_H^{D-3}=\Gamma_{\infty} $
i.e. $R_H$ is defined as the scale at which the small scale
fractal behaviour would match onto the asymptotic density, in
the case that there were a simple cross-over from fractality
to homogeneity. Once $R_H$ has been defined, one can meaningfully 
study correlation properties of fluctuations about the mean density
$\Gamma_\infty$ with the usual normalized correlation 
function $\xi(r)=(\Gamma(r)/\Gamma_{\infty})-1 $.
The relationship of the scale $R_H$ so defined to other characteristic
scales often used is simple to derive. For example, consider the ``correlation
length'' $r_o$ defined by $\xi(r_o)=1$. 
If we assume that $r_o$ lies in the range in which the distribution 
is well approximated as fractal - which is generally taken to be
the case - one finds  $r_o=2^{1/D-3} R_H$. For any particular smooth form
of the cross-over scale from small scale fractality to homogeneity the
 precise relation will be slightly modified.

We now turn to the estimation of the luminosity density from three
dimensional surveys. We adapt here the approximation which is always 
made in this context: We assume that the spatial correlations in
galaxy positions are unconnected to their morphological or luminosity
properties. While such an assumption is known to be  strictly false
(Binggeli et al. 1988) - it is inconsistent with local 
morphological properties (e.g. elliptical galaxies are mostly
located in the center of rich clusters (Dressler 1984), and there is 
a correlation between luminosity and space distributions as discussed
in SLMP98) - 
we will check that it is quantitatively a good approximation for
the estimates being made here. With this assumption we can write
the factorized expression 
\be
\label{e1}
\langle \nu(r,L) \rangle_p dL d^3r  = 
\phi(L)   \Gamma(r)  dL d^3r =
A r^{D-3} L^{\alpha} e^{-\frac{L}{L_*}} d^3r dL
\ee
for the (conditional) average number of galaxies in the volume 
element $d^3r$ at distance $r$ from a observer located on a galaxy,
and with luminosity in the range $[L,L+dL]$.  In the latter form we have 
used the fact that the galaxy luminosity function has been observed 
to have the so-called Schechter shape with parameters
$L_*$ (luminosity cut-off) and $\alpha$ (power law index)
which can be determined experimentally (Binggeli et al. 1988),
and we have written the small scale fractal behavior for the
spatial distribution. 
Hence $\langle \nu(r,L) \rangle_p$ is a function of 
the measurable parameters $L_*, \alpha$  and those
characterizing the spatial distribution - $D, A$ at small
scales and, in the case of detected homogeneity, $R_H$ at
large scales. Note that for the determination of the shape 
of the luminosity function the effect of space inhomogeneities
can be neglected if the joint distribution can be written as in Eq.\ref{e1}
(Binggeli et al. 1988, SLMP98). There 
are different methods to estimate the parameters $M_*$ and $\alpha$
but {\it all} are based on the assumption embodied in Eq.\ref{e1}:
these so-called inhomogeneity-independent methods
have been developed to determine the shape of the luminosity function,
independently of its overall normalization. 
It is now simple to estimate the average luminosity density
in a sphere of radius $R$ and volume $V(R)$ placed
around a galaxy 
\be
\label{e2}
\langle j(<R) \rangle_p = \frac{1}{V(R)} 
\int_0^R \int_{0}^{\infty} \langle \nu(r,L)\rangle_p L dL d^3r 
\ee
which has the $R$ dependence which follows from that of the space
density, with a corresponding asymptotic value in the case
of homogeneity. In a volume limited 
sample (hereafter VL - see e.g. JMSL99) 
extracted from a given 
redshift survey,
we may compute the number of galaxies as a function of distance.
Using Eq. \ref{e1} we have
\be
\label{e3}
\langle N (L>L_{VL}; r <R) \rangle_p 
= \int_0^R \int_{L_{VL}}^{\infty} \langle \nu(r,L) \rangle_p dL d^3r
= B_{VL}R^D
\ee
where $B_{VL}$ is the amplitude of the number counts in a VL sample
with faint luminosity limit at $L=L_{VL}$.
From Eq. \ref{e1} and Eq. \ref{e3} 
and considering Eq.\ref{e2} we then obtain that
\be
\label{e5}
\langle j(<R)  \rangle_p
\equiv j(10) \left(\frac{R}{10h^{-1}}\right)^{D-3}\; ,
\ee
in $L_{\odot} \cdot Mpc^{-3}$, where
$j(10)=
3/(4 \pi)   L_* (10h^{-1})^{D-3} \gamma(\alpha+2) \Phi_{VL}$
where $\gamma(\alpha+2)$ is the Euler function 
($\gamma(n)=(n-1)!$ for positive integers $n$)
\footnote{Note that for the values relevant here $\alpha \approx -1$ 
the integral in the denominator of $\Phi_{VL}$
is a cut-off divergent gamma function ($n<0$), and 
depends sensitively on the lower cut off $y_{{\rm VL}}$. On the other hand
the gamma function  is convergent,  
so that the total luminosity is dominated by galaxies with 
luminosity $\sim L_*$, and is essentially insensitive to the 
lower cut off in the luminosity function $L_{min}$. If there are 
very many additional very low surface brightness  galaxies which 
are not sampled in redshift surveys, sufficient to make the 
exponent $\alpha < -2$, this integral would be strongly dependent 
on $L_{min}$ (and the total luminosity dominated by these 
faint galaxies). }
and we have defined 
$\Phi_{VL} = B_{VL}/ (\int_{y_{VL}}^{\infty} y^{\alpha}e^{-y} dy) \;$
where $y_{VL} = L_{VL}/L_*$.

Employing our definition of the homogeneity scale $R_H$ we obtain
the asymptotic average luminosity density by simply substituting 
$R=R_H$ in Eq.\ref{e5}.  Given a value of (or lower bound on) this
scale it is thus straightforward to obtain the corresponding value 
(or upper bound on) the total mass  density, once one has
an appropriate estimate of the global mass to luminosity ratio.
The numerical results we quote here we obtain from the 
CfA2-South survey (Huchra et al., 1999), 
which covers a solid angle of about one steradian, with a completeness 
in its observing range of over $99 \%$ and a total of 4392 galaxies. 
We have repeated our calculations (Montuori et al. 2001)
in the larger joint catalogue of CfA2 and SSRS2,
including both the Southern and Northern galactic caps, and find 
results which are in good agreement with those given here. 
For the luminosity parameters we take $M_*^{ZW}=-18.8 \pm 0.3$ 
and $\alpha=-1.0 \pm 0.2$  
(Marzke, Huchra \& Geller 1994). Note that
we compute all the relevant quantities in the Zwicky
magnitude system used in these surveys.
The ${\cal M}/{\cal L}$ results from clusters refer to 
luminosity in the $B$ magnitude system, to which the 
transformation from the Zwicky system is not exactly
known (Marzke, Huchra \& Geller 1994). In practice, up
to a small residual effect due to galaxy type, the 
transformation should be well modeled as a simple
zero point  offset $M^{B} = M^{ZW}+\Delta$ with $\Delta \le 0.3$ 
(Paturel et al., 1994). In our estimated luminosity density 
this induces the correction 
$j(10)^B = j(10) \times 10^{-0.4 \Delta}$, which is small,
and we will simply neglect it in what follows.
For the spatial properties the results we quote are for the
analysis of the CfA2 survey described in JMSL99, 
using exactly the methods used there to estimate the appropriate parameters. 
In Table \ref{tab_Bvl} are given the values of $B_{VL}$ 
determined in different VL samples, defined by the corresponding
absolute magnitude limit $M_{VL}$. The results depend  
on the cosmological parameters
assumed in the reconstruction of distances and absolute 
magnitudes from redshifts and apparent magnitude.
The values quoted correspond to the Mattig relation with $q_0=0.5$,
but the results do not sensibly change for any other reasonable
choice of $q_0$ as the redshifts involved are very small ($z \le 0.05$).
From $B_{VL}$ we have computed the quoted values of the quantity
$\Phi_{VL}$ in the different VL samples, and we infer the
average value $\langle \Phi_{VL} \rangle = 1.4 \pm 0.4$.
Using this we obtain the numerical value 
$j(10) \approx (2 \pm 0.6) \times 10^{8} \,   h L_{\odot}/Mpc^3 .
$
The fractal dimension $D$ is given by the slope
of $\langle N(<r) \rangle_p$ as a function of $r$ in a VL sample.
As mentioned above our analysis of the CfA2+SSRS2 joint 
catalogue  gives values very consistent 
with these. 
Hereafter we adopt for simplicity the value $D=2.0$.

For a given $R_H$ we now find the mass density parameter in 
units of the critical density 
$\rho_{c} = 2.78 \cdot 10^{11} h^2 \; M_{\odot}/Mpc^3$
where $M_{\odot}$ is the solar mass, and as a function of 
a specified global mass-to-luminosity ratio 
(in solar  and  $h$ units), to be 
\be
\label{e9}
\Omega_m \left(R_H, \frac{{\cal M}} {{\cal L}}\right) 
= [(6 \pm 2) \times 10^{-4}]  
 \frac{{\cal M}} {{\cal L}} h^{-1}
  \left(\frac{10h^{-1}}{R_H}\right) \;  .
\ee
Note that because  
estimates of ${\cal M}/{\cal L}$ are linearly dependent on  $h$, 
and $R_H$ is measured in units of $h^{-1}$ Mpc, Eq.\ref{e9}
is in fact independent  of the Hubble constant.

Before proceeding to discuss this estimate of $\Omega_m$ in more 
detail we comment on the variation in $\Phi_{VL}$ 
seen in Table \ref{tab_Bvl}. These fluctuations 
can be due to one or more of the following factors:
{\it (i)} Errors (statistical and/or systematic) 
in the measurement of $B_{VL}$:
this effect can be very important for the samples $M_{VL} < M_*$,
as in this range of magnitudes  the statistics of the VL samples
is much weaker because of the exponential break in the 
luminosity function. Further, as discussed in various papers 
(i.e. Bothun \& Cornell 1990),  the magnitudes in the 
GCGC catalogue (from which the photometry of CfA2 comes) are
based on the Zwicky system, with an estimated error
of $0.3^m$ up to $15.0^m$ increasing up to
$\gtapprox 0.5^m$ for the faint  end of the catalogue (i.e. for
$15.0 \ltapprox m \ltapprox 15.7$). The effect of this
systematic error is not crucial in the estimation
of $\Gamma(r)$ if the statistics  is robust (SLMP98), and
in the VL considered there is a good spread
of absolute magnitudes (that is $M_{VL} \gtapprox -19.5$).
Clearly in the deepest and more luminous VL samples
($M_{VL} \ltapprox -20.0$) the effect of the photometry 
error  is   more important in the determination
of the amplitude of the conditional average density,
in view of Malmquist bias (e.g. Teerikorpi 1997).
{\it (ii)} Use of non optimal parameters $\alpha$ and $M_*$
in the computation of $\Phi_{VL}$. In order to check the dependence 
on these two
parameters we have let them vary in the range $-1.2 \le \alpha \le -0.9$
and $-18.7 \le M_* \le -19.1$.
In the VL samples with $M_{VL} > M_*$
we do not see a large fluctuation of $\Phi_{VL}$, while
for the brighter samples it indeed can cause a change by a substantial
factor $10 \% \div 30 \%$. The values we have used here correspond
to $M_*=-19.1$ which give the stablest result for  $\Phi_{VL}$.
{\it (iii)} The breakdown of the assumption of luminosity/space
independence embodied in  Eq.1: The independence of the determined
parameters of the luminosity sample is our consistency test of this
assumption, and to the extent that the fluctuations are relatively small
it is good. Further given the first two points which may explain much of
the observed spread the error caused by this is certainly at most of
the order of the $20 \%$ we have given.

Let us now consider further our estimate of $\Omega_m$.
Taking first the  estimate  ${\cal M}/{\cal L} \approx  10h$ 
in the $B$-band as derived by Faber and Gallagher (1979), 
which corresponds to a global mass to luminosity ratio typical
of spiral galaxies, we obtain
$\Omega_m(R_H) 
\approx  6  \times 10^{-3}\left(\frac{10h^{-1}}{R_H} \right) \; .$
With $R_H \approx 10 \hmp$ ($r_o \approx 5 \hmp$) we obtain
the value $\Omega_g  \approx 6  \cdot 10^{-3}$ of
the standard treatment of Peebles (1993). On the other hand 
we can determine the mass to luminosity ratio which would
give a critical mass density Universe. For given $R_H$ 
we find 
$\left(\frac{{\cal M}}{{\cal L}}\right)_{crit} \approx  1600h  
\left(\frac{R_H}{10h^{-1}}\right) \;, $
so that again the canonically quoted value of 
$({\cal M}/{\cal L})_{crit} \approx  1600h$
corresponds to the homogeneity scale $R_H\approx 10 \hmp$. 

Galaxy clusters have been much studied in recent years,
and they are believed to probe well the global mass to luminosity
ratio, for which the observed value is    
$({\cal M}/{\cal L})_{c} \approx 300h$ in the $B$-band 
(Carlberg et al. 1997, Bahcall 1999, Hradecky et al. 2000).
Taking this value we obtain
$\Omega_m (R_H) \approx  
(0.18 \pm 0.06) \left(\frac{10h^{-1}}{R_H} \right) \; .$
The value which results using the same standard value
$R_H = 10 \hmp$ is $\Omega_m \approx 0.2$ (Bahcall 1999), which
simply means that the former is the homogeneity scale built
into the estimate of the luminosity density from 
Efstathiou et al. (1988) used as the basis for these estimates.
The point of the present paper has been to make the dependence
on this scale explicit, and to use its value as estimated from
three dimensional surveys.   In JMSL99 we placed a lower
bound of $R_H \approx 20 \hmp$ on the homogeneity scale, and 
found no clear statistical evidence for the existence of a cut-off
to homogeneity at larger scales. Using this as 
a {\it conservative lower bound} on
that scale we now obtain the upper bound $\Omega_m \leq 0.1$ on
the total mass density. In SLMP98  a strong
case has been made for a much larger lower bound of $100 \div 150 \hmp$,
based on a combination of cluster catalogues and the LEDA 
red-shift database. While these results remain controversial
(e.g. Wu et al., 2000),
and need confirmation from forthcoming larger red-shift surveys
(2dF and SSDS), it is interesting to consider the implications
of such a finding for determination of $\Omega_m$.

One of the most immediate cosmological implications of the measurement
of the mass density comes from the comparison of its value 
with the standard Big Bang nucleosynthesis (SBBN) limits on the 
baryon density of the universe, which give 
(Olive et al. 2000, Tytler et al. 2000) 
$\Omega_b^{BBN}h^{2} \approx 0.019 \pm 0.004 \;.
$
While this comparison results in the inference of the existence
of non-baryonic dark matter in the standard case, one can now
view it as providing a possible ``window of consistency'' for
the two values. Using the estimate obtained above (\ref{e9})
we find that for  the homogeneity scale 
$R_H  = (0.3 \pm 0.15) {\cal M}/{\cal L}\;{\rm Mpc} $
the dark matter in the Universe can be purely baryonic 
with its global density satisfying the constraints of SBBN. 
Conversely an homogeneity scale larger than this value would
cause a serious problem for the theory of SBBN.  Adopting the 
value  $({\cal M}/{\cal L})_{c} \approx 300h$, we find that 
gives  $R_H = (90 \pm 45) h\;$Mpc which, for  $h=0.65$, 
corresponds to  $R_H = (60 \pm 30)$ Mpc, which allows potential
compatibility even for values of $R_H$ as small as our 
{\it conservative} estimation $R_H=20 \hmp\approx30$Mpc.

Various other methods are commonly used to estimate the mass
density of the Universe. One is based on clusters
is that obtained by observations which constrain the
fraction of hot baryonic X-ray emitting gas to the total
mass in clusters  (see Bahcall 1999). Given that the rest of
the mass may be non-baryonic this gives, when one uses the
nucleosynthesis upper bound on $\Omega_b$, an upper bound on
the total mass $\Omega_m  \le 0.3$. Further, taking most of this mass
to be non-baryonic one infers a value of $\Omega_m$ consistent
with the value from the direct estimate. In the present context
we note simply that, if the scale  $R_H$ is indeed   
larger than usually implicitly assumed, the total mass density 
may be much lower and the dark mass in clusters quite consistently
be baryonic.

Another source of estimates for the total mass density comes
from peculiar velocity flows (see Strauss \& Willick 1994). 
These make use of linear perturbation theory, in which regime one 
can correlate the peculiar motions in a simple way with the total 
mass fluctuations, with the overall amplitude depending on the
total $\Omega_m$ which is then in principle determinable.
In practice the problem is that one does not know how the 
fluctuations in the dominating dark component are related to 
those in the visible matter, and only by making some extremely
simplistic assumptions (e.g. ``linear bias'') can one extract
a result. A much greater problem is one of principle related to
the scale $R_H$, as it is in fact precisely also the scale which
characterizes the validity of a linear regime. If, as we have
discussed here, $R_H$ is much larger than the standard assumed 
value the estimates which have been performed to date are
meaningless. To reliably correlate peculiar velocities with
the mass distribution much tighter constraints are first needed
on the latter, and completely different methods to the standard
ones must be used if the regime of non-linearity extends much
deeper than usually assumed (Joyce et al. 2000).

In this paper we have described, taking the example of the matter 
density $\Omega_m$, how crucial parameters in standard type cosmologies 
are dependent on a scale which has yet to be determined. 
The requisite statistically robust detection and 
characterization of the cross-over to homogeneity and a reliable
determination of this scale will be possible with the forthcoming 2dF 
and SSDS surveys. On the basis of current data we have
placed constraints on $\Omega_m$
by using 
the most suitable publically available three dimensional data 
available for this purpose, the combined CfA2 and SSRS2 surveys. 
We note that we have assumed, as is usually done, 
that clusters do indeed give a reliable measure of the global mass to
luminosity ratio, and that if this assumption is not correct our results
for the estimated parameters will of course not hold.

We warmly thank Y.V. Baryshev, F. Combes,
R. Durrer, P. Ferreira, A. Gabrielli, M. Montuori, 
D. Pfenniger and L. Pietronero for very useful 
comments and discussions.  
F.S.L. acknowledges the support of the  
EC TMR Network  ``Fractal structures and   
self-organization''   
\mbox{ERBFMRXCT980183} and of the Swiss NSF.

\clearpage 

\begin{table}
 \caption{\label{tab_Bvl} Values of $B_{VL}$ 
estimated in volume limited
samples 
in the CfA2 South catalogue. 
The absolute magnitude cut is at $M_{VL}^{ZW}$,
In the third column is shown 
the determined value of the parameter $\Phi_{VL}$
where we have used $M_*=-19.1$ and $\alpha=-1.00$
as parameters of the luminosity function.
$N_{VL}$ is the number of points in the volume-limited sample.}
 \begin{center}
 \begin{tabular}{|c|c|c|c|}
 \hline
          &                   &                        &           \\
$M_{VL}^{ZW}$  & $B_{VL}$     & $\Phi_{VL}$            & $N_{VL}$  \\
          &                   &                        &           \\
\hline		  
 -17.0    & 1.5 $\pm 0.1$     &  1.0 $\pm 0.1$       &  1641      \\
          &                   &                        &           \\
 -18.0    & 0.7 $\pm 0.1$     &  0.9 $\pm 0.1$       &  2518     \\
          &                   &                        &           \\
 -19.0    & 0.4 $\pm 0.05$    &  1.6 $\pm 0.2$       &  4134     \\
          &                   &                        &           \\
 -19.5    & 0.17 $\pm 0.02$   &  1.5 $\pm 0.2$       &  3868     \\
          &                   &                        &           \\
 -20.0    & 0.06 $\pm 0.01 $  &  1.8 $\pm 0.3$       &  2524     \\
          &                   &                        &           \\
\hline
\end{tabular}
\end{center}
\end{table}

\begin{thebibliography}{}
 
 \bibitem{bah99}
Bahcall N., 1999
In the Proc. of the Conference 
``Particle Physics and the Universe (astro-ph/9901076)




\bibitem{bin88}  Binggeli, B., Sandage, A.,
Tammann, G. A. 1998, 
 Astron. Astrophys. Ann. Rev.  26,  509   

\bibitem{bc90} Bothun G.D. and Cornell M.E. 1990
Astron.J. 99,  1004, 
 

 \bibitem{galclu} Carlberg R.G., Yee H.K and Ellingson E., 1997
 Astrophys. J. 478, 462 


 
\bibitem{dre84}
 Dressler A.,1984,  Ann.Rev. Astron.Astrophys.   313,  42


\bibitem{lumfun} Efstathiou G, Ellis R.S. and Peterson B., 1988
Mon.Not.R.astr.Soc  232, 431

\bibitem{fg79} Faber S.M. \& Gallagher J.S., 1979,
Astron.Astrophys. Ann.Rev.  17,  135 
 
\bibitem{gsl01} 
Gabrielli A. and Sylos Labini F., 2001, Europhys.Lett., 
54, 1 

\bibitem{hra00} Hradecky V., Jones C.,  Donnelly R.H.,
Djorgovski S.G., Gal R. R., Odewahn S.C.
2000, Astrophys.J. 543, 521

\bibitem{huchra99} Huchra J.P., Vogeley M.S. and Geller M.J., 1999
 Astrophys.J. Suppl.121, 287 

\bibitem{jmsl99} Joyce M., Montuori M., Sylos Labini F. 1999,
  Astrophys. Journal Letters,  514, L5 (JMSL99)

\bibitem{jampsl00} Joyce M., Anderson P.W., Montuori M.,  
Pietronero L. and Sylos Labini F. 2000,
Europhys.Letters 50, 416 
 
\bibitem{mon00} 
Montuori M. et al., in preparation (2001)
 
\bibitem{olive-nucleo2000}
Olive K.,Steigman G. and Walker T.,
2000, Phys.Rept. 333, 389 

\bibitem{paturel94}
Paturel G., Bottinelli L. and Gouguenheim L., 1994
Astron.Astrophys.  268, 768

\bibitem{pee93} Peebles, P.E.J.,    
``Principles of Physical Cosmology'', Princeton Univ. Press, 1993  
 
\bibitem{saslaw2000} 
Saslaw W.C. ``The distribution of the Galaxies'', Cambridge University
Press 2000

\bibitem{sw94}  
Strauss M. and Willick J., 1995, 
  Physics Reports   261, 271  


\bibitem{slmp98} Sylos Labini F., Montuori M.,  
 Pietronero L., 1998,   Phys.Rep.   293, 66   (SLMP98)

\bibitem{tee97} 
Teerikorpi, P., 1997,  Ann.Rev.Astron.Astrophys 35, 101

\bibitem{tytler-nucleo2000}
Tytler D. et al., astro-ph/0001318, to appear in Physica Scripta.

\bibitem{rees99}   Wu K.K., Lahav O.and Rees M. 
1999, Nature,   225, 230 

\end{thebibliography}
\end{document}